\documentclass{article}
\usepackage{float}

\usepackage{graphicx}  
\usepackage{color}
\usepackage{amsmath,amssymb}  
\usepackage{mathtools}
\mathtoolsset{showonlyrefs,showmanualtags} 
\usepackage{natbib} 
\usepackage{longtable}
\usepackage{lscape}
\usepackage{enumerate}
\usepackage[toc,page]{appendix}
\usepackage{arydshln}
\usepackage{threeparttable} 

\usepackage{authblk}
\usepackage{booktabs}

\usepackage[left=2cm,top=2cm,right=2cm,bottom=2cm,nohead]{geometry}
\newcommand{\be}{\begin{enumerate}}
\newcommand{\ee}{\end{enumerate}}
\newcommand{\bfr}{\begin{frame}}
\newcommand{\efr}{\end{frame}}
\newcommand{\bi}{\begin{itemize}}
\newcommand{\ei}{\end{itemize}}

\def\sym#1{\ifmmode^{#1}\else\(^{#1}\)\fi}

\title{Returns and Order Flow Imbalances: Intraday Dynamics and Macroeconomic News Effects\thanks{This paper is based on part of my PhD dissertation at the Kellogg School of Management, Northwestern University. I am indebted to my dissertation committee—Torben Andersen, Viktor Todorov, Robert Korajczyk, and Beth Andrews—for their invaluable guidance. 
I am also grateful to Nikolaus Hautsch, Takaki Hayashi, Yasuhiro Omori, Kosuke Oya, Ernst Schaumburg, Toshiaki Watanabe, and Tomasz Wisniewski for their helpful comments and suggestions. 
Financial support from the Kellogg School of Management, Osaka University, and Hosei University is gratefully acknowledged. This research was also partially supported by JSPS KAKENHI (A26245028a, 16H036050, T15K170370, and 23K01338), the Joint Usage and Research Center at the Institute of Economic Research, Hitotsubashi University, and research projects of Hosei University’s Innovation Management Research Center (“Statistical Analysis on Information Propagation in Financial Markets and Related Areas” and “Market Quality Analysis Using High-Frequency Order Data”). 
All remaining errors are my own.}
}
\author{Makoto Takahashi\thanks{Email: \texttt{m-takahashi@hosei.ac.jp}}}
\affil{Faculty of Business Administration, Hosei University, Tokyo, Japan}
\date{\empty}

\begin{document}

\maketitle

\begin{abstract}
We study the interaction between returns and order flow imbalances in the S\&P 500 E-mini futures market using a structural VAR model identified through heteroskedasticity. The model is estimated at one-second frequency for each 15-minute interval, capturing both intraday variation and endogeneity due to time aggregation.
We find that macroeconomic news announcements sharply reshape price–flow dynamics: price impact rises, flow impact declines, return volatility spikes, and flow volatility falls. Pooling across days, both price and flow impacts are significant at the one-second horizon, with estimates broadly consistent with stylized limit-order-book predictions. Impulse responses indicate that shocks dissipate almost entirely within a second. Structural parameters and volatilities also exhibit pronounced intraday variation tied to liquidity, trading intensity, and spreads.
These results provide new evidence on high-frequency price formation and liquidity, highlighting the role of public information and order submission in shaping market quality.
\end{abstract}

\noindent%

\section{Introduction}

Modern electronic markets operate on a limit order book (LOB), a dynamic repository of buy and sell quotes at various price levels.\footnote{Most financial markets, including major exchanges such as NASDAQ, NYSE, and Euronext, employ electronic LOB systems.} The LOB continuously evolves as new limit orders, market orders, and cancellations arrive. These order flows generate price impact—the immediate change in prices triggered by incoming orders—which reflects the state of market liquidity. Conversely, price changes can in turn trigger further order submissions or cancellations when traders employ price-contingent strategies. Understanding this two-way interaction between order flows and price changes is fundamental for explaining the price formation mechanism in modern markets.

The market microstructure literature has examined many dimensions of price impact, from the role of trade size and direction to the influence of bid–ask spreads and information asymmetry. A large empirical body also documents intraday patterns in liquidity and trading intensity. While these studies provide valuable insights, they often analyze endogeneity and intraday variation separately rather than within a unified framework. A more comprehensive approach is needed to fully capture how prices and order flows interact in real time. A detailed review of related studies is provided in Section~\ref{sec:literature}.

A closely related line of research is \citet{Fleming2018}, who analyze the U.S.\ Treasury interdealer market on BrokerTec using a vecotre autoregressive (VAR) model of prices and order flow. They show that both trades and limit orders affect prices, and that the variation in limit-order activity makes its contribution to price-update variance especially large; further, macroeconomic announcements amplify these impacts, particularly for limit orders. Our approach complements and extends this evidence. We study a different market—equity index futures—and focus on order flow imbalance (OFI) at the best bid and offer, constructed from one-second BBO data for the S\&P~500 E-mini contract. Rather than imposing a recursive VAR, we employ a bivariate structural VAR identified through heteroskedasticity (SVAR–ITH), which allows us to disentangle the contemporaneous price impact of OFI from the reverse flow impact of returns. By re-estimating the system across short intraday intervals, we trace the within-day dynamics of these impacts and show how they respond to macroeconomic news releases.

Our contributions are threefold.  
First, we provide an integrated framework that simultaneously addresses endogeneity and intraday variation—two central features of high-frequency markets that prior studies have largely examined separately.  
Second, we document pronounced within-day variation in both price and flow impacts, linking them to liquidity conditions and order submission behavior.  
Third, we demonstrate that macroeconomic news announcements systematically reshape the return–flow relationship, temporarily raising price impact and dampening flow impact, with implications for trading strategies and liquidity provision.

The remainder of the paper is organized as follows. 
Section~\ref{sec:literature} reviews related literature. 
Section~\ref{sec:data} describes the dataset and variable construction. 
Section~\ref{sec:methodology} presents the SVAR framework and estimation methodology. 
Section~\ref{sec:empirical-results} reports the empirical findings. 
Finally, Section~\ref{sec:conclusion} concludes.

\section{Related Literature}
\label{sec:literature}

The interaction between order flows and price changes has been a central theme in market microstructure research, spanning several strands of the literature.

A first strand examines the price impact of trades and orders. Since \citet{Hasbrouck1991a}, VAR models have been widely used to show that trades convey information and exert a lasting influence on prices. This empirical tradition builds on asymmetric information models such as \citet{Bagehot1971}, \citet{Kyle1985}, and \citet{Glosten1985}.

A second strand focuses on the structure of the LOB. \citet{Bouchaud2002, Bouchaud2004, Bouchaud2006} documented stylized facts such as the concavity of the price impact function and the long memory of order signs. \citet{Cont2014} developed a stylized model linking order flow imbalance to short-horizon price changes, and showed that estimated price impact varies systematically across the day.\footnote{Market activity is well known to follow a U-shaped intraday pattern, with high activity at the open, a midday trough, and renewed activity into the close; see \citet{wood1985}, \citet{McInish1992}, and \citet{Andersen1997}.\label{footnote-intraday-pattern}}

A third strand highlights the simultaneity of prices and order flows. Because traders often employ price-contingent strategies, the two are jointly determined. \citet{Deuskar2011} addressed this endogeneity using the ITH approach of \citet{Rigobon2003} and its extensions, showing that ignoring simultaneity biases price impact estimates and masks important variation in liquidity conditions.

Our study contributes to all three strands by combining insights from the VAR tradition, the LOB literature, and ITH-based identification into a unified framework. Using one-second data, we explicitly capture both the intraday variation in market activity and the bidirectional causality between prices and order flows.

\section{Data and Variable Construction}
\label{sec:data}

\subsection{Market and Dataset Overview}
The dataset used in this study is constructed from the BBO files of the S\&P 500 E-mini futures contract obtained from CME Group.\footnote{The E-mini is a liquid asset and most of the price discovery for the S\&P 500 occurs in the E-mini market (e.g., \cite{Hasbrouck2003}). The minimum contract size is \$50 times the futures price and the minimum tick size is 0.25 index points, which is equivalent to \$12.5 ($=50 \times 0.25$). It is traded nearly 24 hours from Monday to Friday. Regular trading starts at 8:30 Chicago Time (daylight savings time) and ends at 15:15. After a 15-minute break, electronic trading is available from 15:30 to 8:30 except for a 30-minute daily maintenance shutdown from 16:30. Contract months are March, June, September, and December and are available with the latest five months in the March quarterly cycle. Trading can occur up to 8:30 on the third Friday of the contract month.} 
The BBO file records all events that change the price or size of the best bid and ask quotes, including transactions from marketable orders and quote revisions from limit orders and cancellations. Each event is time-stamped to the second with a unique sequence number for the day.

The sample covers 1,490 trading days from January 2, 2008 to December 31, 2013.\footnote{Days with insufficient observations—mostly before/after holidays—are excluded.} For each day, we extract observations from 8:30–15:00 Chicago Time, corresponding to NYSE regular trading hours when most S\&P 500 components trade. To ensure data consistency, we select the most active contract (highest daily trading volume) each day.\footnote{The most active contract typically rolls from the front month to the next about a week before the expiry.}

\subsection{Variable Construction}
From the BBO data, mid-quote returns, denoted by $r_{t}$, are computed every second. Order flow imbalances (OFI) are constructed following \cite{Cont2014}. Let $P_{n}^{a}$ and $q_{n}^{a}$ denote the best ask price and size, and $P_{n}^{b}$ and $q_{n}^{b}$ the best bid price and size. The order book event is defined as:
\[
e_{n} = q_{n}^{b} I_{\{P_{n}^{b} \geq P_{n-1}^{b}\}} - q_{n-1}^{b} I_{\{P_{n}^{b} \leq P_{n-1}^{b}\}} - q_{n}^{a} I_{\{P_{n}^{a} \leq P_{n-1}^{a}\}} + q_{n-1}^{a} I_{\{P_{n}^{a} \geq P_{n-1}^{a}\}},
\]
where $I_{\{A\}}$ is an indicator function.  
Aggregating $e_{n}$ over one-second intervals yields the order flow imbalance $f_{t}$.\footnote{If order types are available, OFI can alternatively be computed by summing market, limit, and cancellation orders on both sides.}

We also compute additional market activity measures:  
\begin{itemize}
\item Number of order book events and average size of events (order activity and aggressiveness)  
\item Average spread (transaction cost proxy)  
\item Depth, calculated as the average of bid and ask sizes around price changes:
\begin{align}
D_{t}
= \frac{1}{2} \left[ \frac{\sum_{n \in I_{t}} ( q_{n}^{b} I_{\{P_{n}^{b} < P_{n-1}^{b}\}} + q_{n-1}^{b} I_{\{P_{n}^{b} > P_{n-1}^{b}\}} )}{\sum_{n \in I_{t}} I_{\{P_{n}^{b} \neq P_{n-1}^{b}\}}} + \frac{\sum_{n \in I_{t}} ( q_{n}^{a} I_{\{P_{n}^{a} > P_{n-1}^{a}\}} + q_{n-1}^{a} I_{\{P_{n}^{a} < P_{n-1}^{a}\}} )}{\sum_{n \in I_{t}} I_{\{P_{n}^{a} \neq P_{n-1}^{a}\}}} \right], \label{eq-depth}
\end{align}
where $I_{t}$ represents a one-second interval.
\end{itemize}

\subsection{Summary Statistics}
Table \ref{tab-sumstat} presents summary statistics for mid-quote returns, OFI, and market activity variables.  
Both returns and OFI are roughly symmetric around zero but highly variable. OFI percentiles reveal periods with intense one-sided order submission. On average, 45 events occur per second, each averaging 15 contracts. While the spread remains at the minimum tick size (0.25), depth is substantial compared with event size.

\begin{table}[tbp]\centering
\begin{threeparttable}
\def\sym#1{\ifmmode^{#1}\else\(^{#1}\)\fi}
\caption{Summary statistics of mid-quote returns, OFI, and market activity variables}\label{tab-sumstat}
{\small
\begin{tabular}{lrrrrrrrrr}
\addlinespace
\toprule
& \multicolumn{1}{c}{Mean} &  \multicolumn{1}{c}{SD} &  \multicolumn{1}{c}{1\%} &  \multicolumn{1}{c}{5\%} &  \multicolumn{1}{c}{25\%} &  \multicolumn{1}{c}{50\%} &  \multicolumn{1}{c}{75\%} &  \multicolumn{1}{c}{95\%} &  \multicolumn{1}{c}{99\%} \\
\midrule
Mid-Quote Return\sym{*} & 0.00 & 0.91 & $-2.86$ & $-1.78$ & 0.00 & 0.00 & 0.00 & 1.78 & 2.87 \\
\addlinespace
Order Flow Imbalance\sym{***} & $-0.00$ & 0.52 & $-1.60$ & $-0.48$ & -0.03 & 0.00 & 0.03 & 0.47 & 1.59 \\
\addlinespace
Number of Events\sym{**} & 0.45 & 0.66 & 0.01 & 0.03 & 0.10 & 0.23 & 0.51 & 1.72 & 3.27 \\
\addlinespace
Average Size of Events\sym{**} & 0.15 & 0.36 & 0.01 & 0.01 & 0.03 & 0.06 & 0.13 & 0.59 & 1.43 \\
\addlinespace
Average Spread & 0.25 & 0.01 & 0.25 & 0.25 & 0.25 & 0.25 & 0.25 & 0.26 & 0.30 \\
\addlinespace
Depth\sym{***} & 0.63 & 0.43 & 0.07 & 0.12 & 0.33 & 0.55 & 0.84 & 1.43 & 2.05 \\
\bottomrule
\end{tabular}
}
\begin{tablenotes}
\footnotesize
\item \textit{Notes:} Statistics are computed from 34,512,298 samples of returns, flow, and the number and average size of events, and from 5,874,016 samples of depth, all measured at one-second intervals. Depth is defined in Eq.~\eqref{eq-depth}. \sym{*} in basis points, \sym{**} in hundreds, and \sym{***} in thousands.
\end{tablenotes}
\end{threeparttable}
\end{table}

These statistics confirm that the E-mini market is highly liquid and active, with multiple events per second—raising potential endogeneity concerns if returns and flows are aggregated.

\subsection{Intraday Variations}
\begin{figure}[htbp]
\begin{center}
\begin{tabular}{cc}
Standard Deviation of Returns & Standard Deviation of Order Flows \\
\includegraphics[width=.47\linewidth]{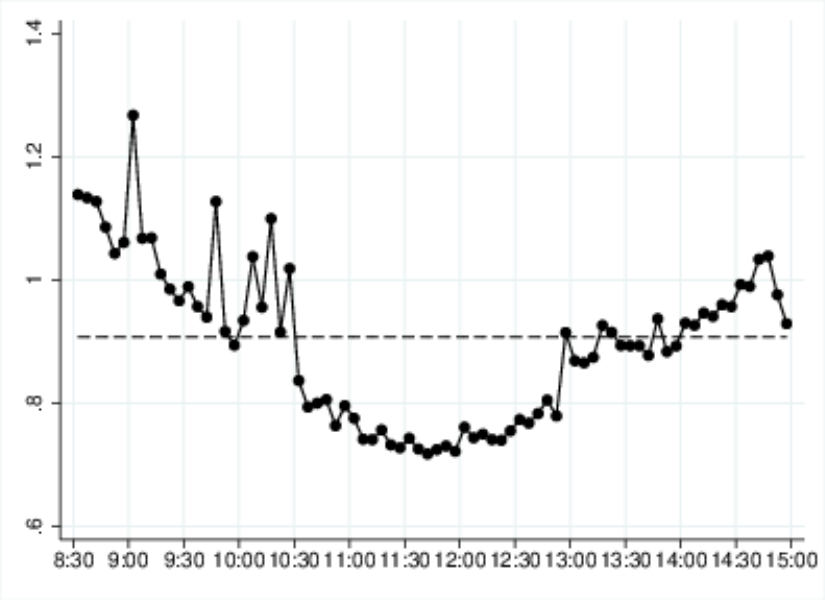} & \includegraphics[width=.47\linewidth]{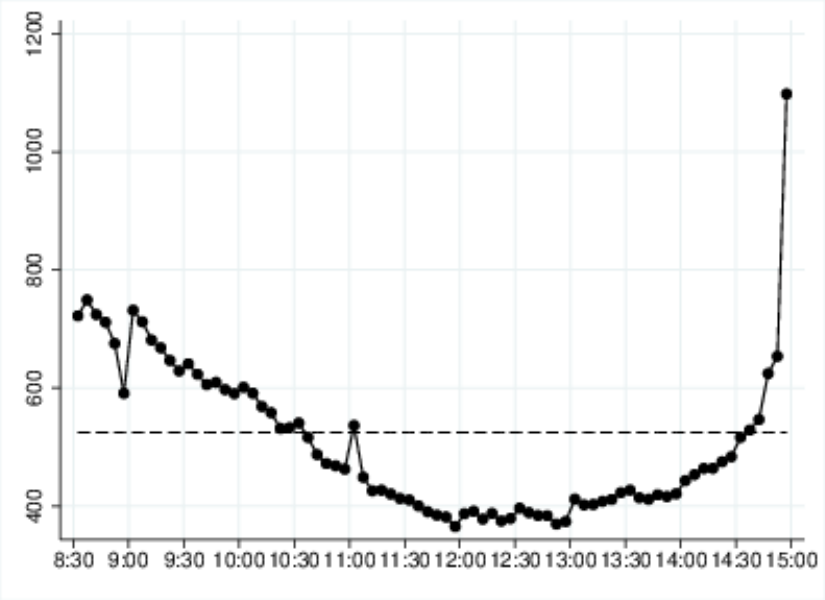}
\\
Mean of the Number of Events & Mean of the Average Size of Events \\
\includegraphics[width=.47\linewidth]{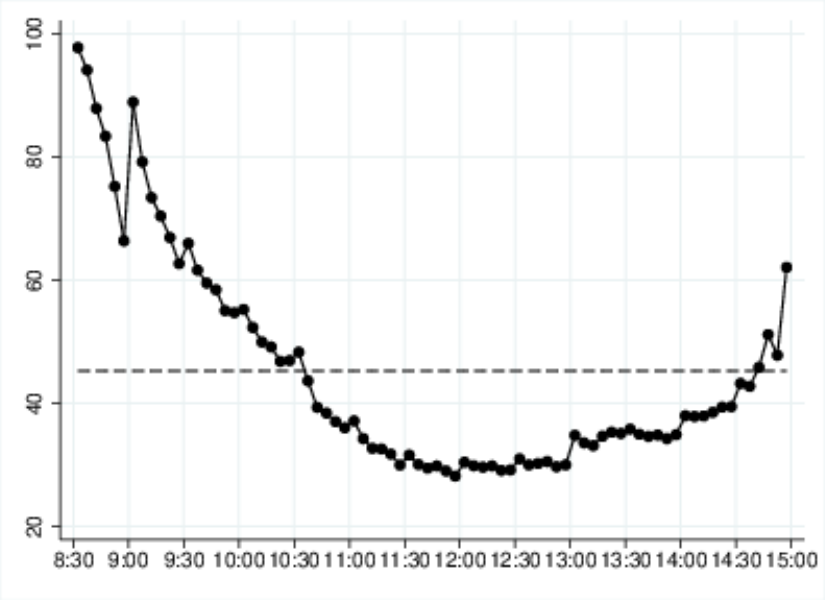} & \includegraphics[width=.47\linewidth]{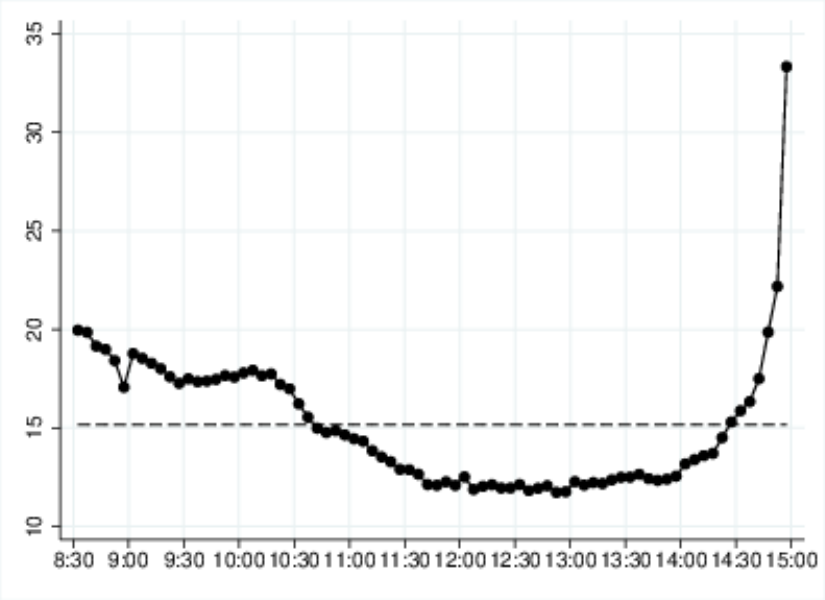}
\\
Mean of the Average Spread & Mean of the Depth \\
\includegraphics[width=.47\linewidth]{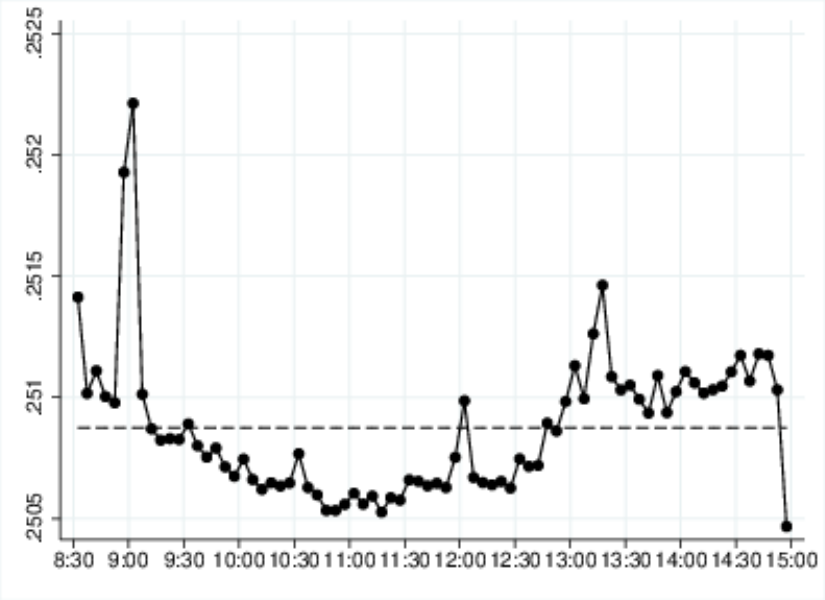} & \includegraphics[width=.47\linewidth]{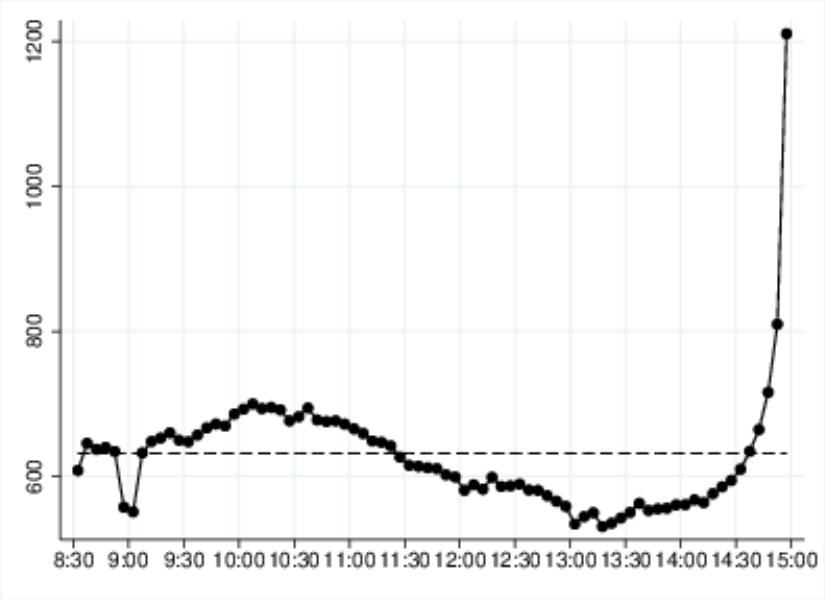}
\end{tabular}
\caption{Standard deviations of mid-quote returns and order flow imbalances, and means of the number of events, average event size, average spread, and depth. Statistics are computed for each five-minute interval between 8:30 and 15:00. Dashed lines indicate the overall averages across all samples.}
\label{fig-intra-var}
\end{center}
\end{figure}

Figure \ref{fig-intra-var} illustrates the intraday patterns of return and OFI volatilities, as well as various market activity measures.
The standard deviations of returns and OFI, as well as the number and size of events, exhibit the familiar U-shaped pattern, with heightened activity at the open and close.\footnote{See footnote~\ref{footnote-intraday-pattern} for prior literature on U- and J-shaped intraday patterns.}  

Notably:
\begin{itemize}
\item In the last 5 minutes (14:55–15:00), OFI volatility, event size, and depth rise sharply, while return volatility and spread decline—consistent with high liquidity.
\item Around 9:00, return volatility and spreads spike while order activity drops, reflecting temporary illiquidity.
\end{itemize}

These patterns underscore the importance of accounting for intraday variation when modeling the return–order flow interaction (Section \ref{sec:methodology}). From a practical perspective, these shifts in market conditions imply that high-frequency traders, liquidity providers, and algorithmic strategies face very different execution environments across the day. Ignoring such variation can lead to misestimation of price impact and potentially suboptimal trading or risk management decisions.

\section{Methodology}
\label{sec:methodology}

This section outlines the model and estimation framework used to examine the interaction between asset returns and order flow imbalances. The analysis builds on the simultaneous equations framework of \citet{Deuskar2011}. First, a simple bivariate model is introduced to illustrate the endogeneity problem. Second, the ITH method is described under a simplified setting. Finally, a SVAR model is presented, which incorporates both endogeneity and the dynamic interaction between returns and flows, and we define measures of instantaneous and long-run impacts. For brevity, the term “order flows” is used interchangeably with “order flow imbalances” in what follows.

\subsection{Simple Bivariate Model}
The contemporaneous impact of order flows on returns is captured by the following equation:
\begin{align}
r_{t} = b_{r} f_{t} + \epsilon_{r,t}, \qquad \epsilon_{r,t} \sim (0,\omega_{r}^{2}), \label{eq-return}
\end{align}
where $r_{t}$ and $f_{t}$ denote returns and order flows, respectively, and $\epsilon_{r,t}$ is a return innovation with mean zero and variance $\omega_{r}^{2}$. The coefficient $b_{r}$ is interpreted as the \emph{price impact} of order flows, reflecting market illiquidity. A larger $b_{r}$ implies greater sensitivity of returns to order flows and therefore lower liquidity, whereas $b_{r}=0$ indicates a perfectly liquid market.

At ultra-high frequencies, the price impact of an individual order is largely mechanical, depending on the depth of the LOB. Under the stylized LOB model of \citet{Cont2014}, the price impact is approximately $1/2D$, where $D$ is the market depth. Using NYSE Trades and Quotes data, \citet{Cont2014} verify that the price impact of aggregated order flow imbalances is consistent with this stylized model.

In practice, however, returns and flows are measured over short intervals that contain many orders. This aggregation introduces \emph{endogeneity}, since price changes can trigger further order submissions within the same interval. \citet{Deuskar2011} trace this simultaneity to stale limit orders and price-contingent trading strategies discussed by \citet{Obizhaeva2008} and \citet{Obizhaeva2013}. Accounting for this endogeneity requires an additional equation for order flows:
\begin{align}
f_{t} = b_{f} r_{t} + \epsilon_{f,t}, \qquad \epsilon_{f,t} \sim (0,\omega_{f}^{2}), \label{eq-flow}
\end{align}
where $\epsilon_{f,t}$ is an innovation in flows, with mean zero and variance $\omega_{f}^{2}$, assumed uncorrelated with $\epsilon_{r,s}$ for any $s$. The coefficient $b_{f}$ measures the \emph{flow impact}, i.e., the feedback from returns to concurrent order flows, and should be zero if no endogeneity exists.

The two equations can be summarized in matrix form:
\begin{align}
B y_{t} = \epsilon_{t}, \qquad \epsilon_{t} \sim (0, \Omega), \label{eq-simul}
\end{align}
where $y_{t} = (r_{t}, f_{t})'$, $\epsilon_{t} = (\epsilon_{r,t}, \epsilon_{f,t})'$, and
\begin{align}
B = \begin{pmatrix} 1 & -b_{r} \\ -b_{f} & 1 \end{pmatrix}, \qquad
\Omega = \begin{pmatrix} \omega_{r}^{2} & 0 \\ 0 & \omega_{f}^{2} \end{pmatrix}. \label{eq-B-Omega}
\end{align}
This implies $y_{t} = B^{-1} \epsilon_{t}$, where $B^{-1}$ is non-diagonal, meaning $f_{t}$ and $\epsilon_{r,t}$ are correlated. As a result, OLS applied to Eq.~\eqref{eq-return} produces an inconsistent estimator of $b_{r}$ when $b_{f} \neq 0$. To obtain consistent estimates, we adopt the ITH method described below.

\subsection{ITH Method}
Let $\Sigma$ denote the variance–covariance matrix of $y_{t}$:
\begin{align}
\Sigma = \text{Var}(y_{t}) = 
\begin{pmatrix} 
\sigma_{r}^{2} & \sigma_{rf} \\ 
\sigma_{rf} & \sigma_{f}^{2} 
\end{pmatrix}. \label{eq-Sigma}
\end{align}
From Eq.~\eqref{eq-simul}, the relationship $B \Sigma B' = \Omega$ holds, which yields three equations:
\begin{align}
& \sigma_{r}^{2} - 2 b_{r} \sigma_{rf} + b_{r}^{2} \sigma_{f}^{2} - \omega_{r}^{2} = 0, \label{eq-cov-1} \\
& \sigma_{f}^{2} - 2 b_{f} \sigma_{rf} + b_{f}^{2} \sigma_{r}^{2} - \omega_{f}^{2} = 0, \label{eq-cov-2} \\
& b_{f} \sigma_{r}^{2} - (1 + b_{r} b_{f}) \sigma_{rf} + b_{r} \sigma_{f}^{2} = 0. \label{eq-cov-3} 
\end{align}
The sample variances and covariance provide estimates of $\sigma_{r}^{2}$, $\sigma_{f}^{2}$, and $\sigma_{rf}$. However, the system contains four unknowns $(b_{r}, b_{f}, \omega_{r}, \omega_{f})$ but only three equations, so identification fails without further restrictions.

The ITH method resolves this by exploiting heteroskedasticity. Suppose there are $S$ states such that the variance–covariance matrix $\Sigma$ changes across states, while $b_{r}$ and $b_{f}$ in $B$ remain constant. For each state $s=1,\dots,S$, Eqs.~\eqref{eq-cov-1}--\eqref{eq-cov-3} hold, with $\omega_{r,s}$ and $\omega_{f,s}$ varying by state. 

With $S$ states, there are $3S$ equations and $2 + 2S$ unknowns. If $S \ge 2$ (order condition) and the following rank condition holds for any $s' \neq s''$,
\begin{align}
\sigma_{r,s'}^{2}\sigma_{rf,s''} - \sigma_{r,s''}^{2}\sigma_{rf,s'} \neq 0, \label{eq:rank}
\end{align}
the parameters are identified.\footnote{See \citet{Rigobon2003} for details on estimation and identification in general settings.} Under these conditions, the parameters can be estimated by the generalized method of moments (GMM) using the $3S$ moment conditions implied by Eqs.~\eqref{eq-cov-1}--\eqref{eq-cov-3}.

\citet{Deuskar2011} estimate this bivariate model assuming $b_{r}$ and $b_{f}$ are constant, while $\omega_{r}^{2}$ and $\omega_{f}^{2}$ vary across the day. They find both $b_{r}$ and $b_{f}$ are significantly positive, confirming endogeneity in flows and the resulting simultaneity bias in price impact estimates.

\subsection{SVAR Model}
The simple bivariate model \eqref{eq-simul} addresses endogeneity but ignores the serial dependence of returns and flows. To account for both endogeneity and dynamic interaction, we extend it to a SVAR:
\begin{align}
B y_{t} = c + \Phi_{1} y_{t-1} + \Phi_{2} y_{t-2} + \cdots + \Phi_{p} y_{t-p} + \epsilon_{t}, \qquad \epsilon_{t} \sim (0, \Omega), \label{eq:SVAR}
\end{align}
where $B$ and $\Omega$ are as in Eq.~\eqref{eq-B-Omega} and $\Phi_{j}$ are $2 \times 2$ matrices capturing autocorrelation.

The SVAR can be rewritten in reduced form:
\begin{align}
y_{t} = \tilde{c} + \tilde{\Phi}_{1} y_{t-1} + \tilde{\Phi}_{2} y_{t-2} + \cdots + \tilde{\Phi}_{p} y_{t-p} + \eta_{t}, \label{eq:reduced-form-VAR}
\end{align}
where $\tilde{c} = B^{-1} c$, $\tilde{\Phi}_{j} = B^{-1} \Phi_{j}$, and $\eta_{t} = B^{-1} \epsilon_{t}$.
The residuals $\hat{\eta}_{t}$ from the reduced-form VAR provide inputs to the ITH estimation, which yields estimates of $b_{r}$, $b_{f}$, and the standard deviations $\omega_{r}$ and $\omega_{f}$.

Once the SVAR is identified, impulse response functions (IRFs) trace the effects of shocks. Let $IRF_{ij}(k)$ denote the response of variable $i$ to a one-unit shock in innovation $j$ at horizon $k$, where $i,j \in \{ r, f \}$. For example, $IRF_{rf}(k)$ measures the impact of order flow shocks on returns at lag $k$. The cumulative response up to horizon $K$ is:
\[
I_{ij}(K) = \sum_{k=0}^{K} IRF_{ij}(k).
\]

The \emph{instantaneous impact} is $I_{ij}(0) = IRF_{ij}(0)$. The \emph{long-run impact} is defined as
\begin{align}
I_{ij}(\infty) = \sum_{k=0}^{\infty} IRF_{ij}(k) = \left[(I_{2} - \tilde{\Phi}_{1} - \tilde{\Phi}_{2} - \cdots - \tilde{\Phi}_{p})^{-1} B^{-1}\right]_{ij}, \label{eq-irf-inf}
\end{align}
where $I_{2}$ is the $2 \times 2$ identity matrix and $[A]_{ij}$ denotes the $(i,j)$ element of matrix $A$.

\section{Empirical Results}
\label{sec:empirical-results}

This section presents the main empirical findings from the SVAR model estimated using one-second BBO data for the S\&P 500 E-mini futures contract. The model is estimated separately for each 15-minute interval of each trading day, allowing us to capture time variation in structural parameters and impulse responses. We first highlight the effects of macroeconomic news announcements, as they represent the most pronounced source of variation in the return–flow relationship. We then present the estimation results for structural parameters and innovation volatilities, followed by impulse response functions, and finally discuss intraday patterns.

\subsection{Macroeconomic News Effects}\label{sec:macro-news}

Macroeconomic announcements are a key source of public information that sharply alters trading activity, liquidity, and the relationship between prices and order flows. Prior research such as \citet{Andersen2003b, Andersen2007c} and \citet{Hautsch2011b} shows that these announcements trigger intense trading and spikes in return volatility, while \citet{Brogaard2014} demonstrate that high-frequency traders actively adjust their strategies in response. Building on this literature, we investigate how major U.S. macroeconomic releases shape the structural parameters estimated in our SVAR model.

We focus on the eight announcements analyzed by \citet{Brogaard2014}: construction spending, consumer confidence, existing home sales, factory orders, ISM manufacturing, ISM services, leading indicators, and wholesale inventories. Most of these announcements are released at 9:00 Chicago Time.\footnote{Exceptions include ISM services (7:55 on February 5, 2008 and 10:00 on August 5, 2008) and consumer confidence (8:37 on June 28, 2011).} 

\begin{figure}[tbp]
\centering
\begin{tabular}{cc}
$b_{r}$ & $b_{f}$ \\
\includegraphics[width=.47\linewidth]{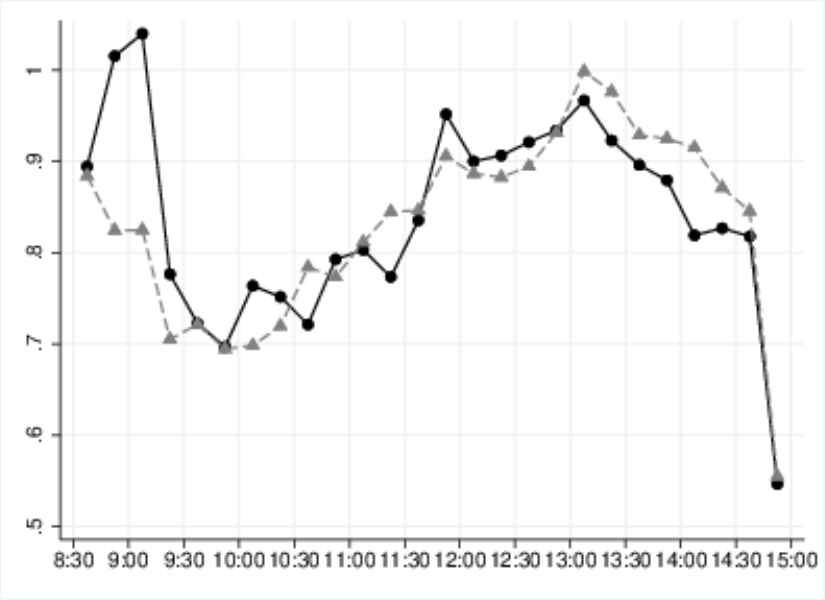} & \includegraphics[width=.47\linewidth]{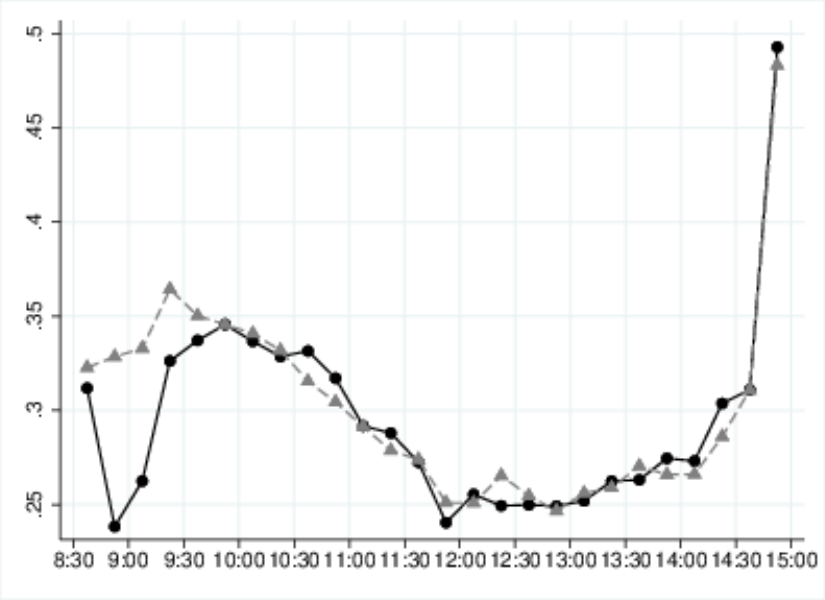} \\
$\omega_{r}$ & $\omega_{f}$ \\
\includegraphics[width=.47\linewidth]{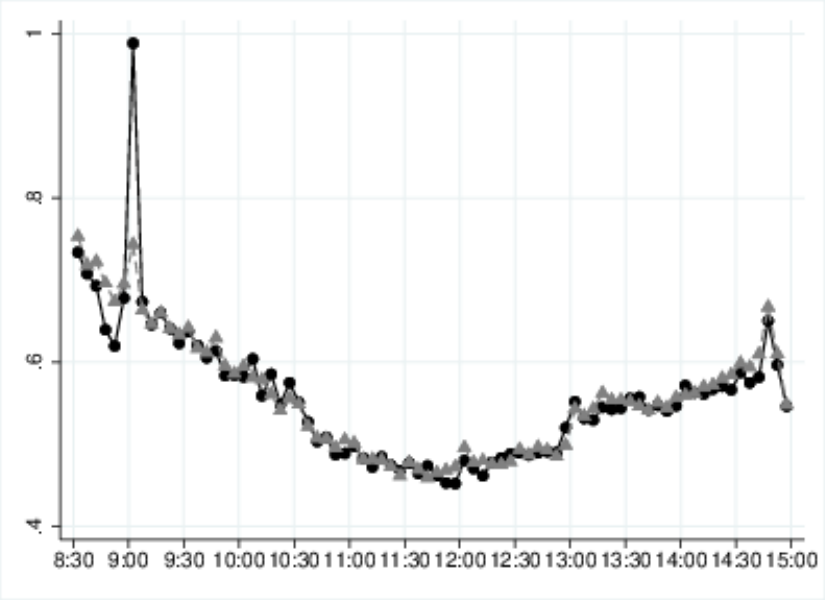} & \includegraphics[width=.47\linewidth]{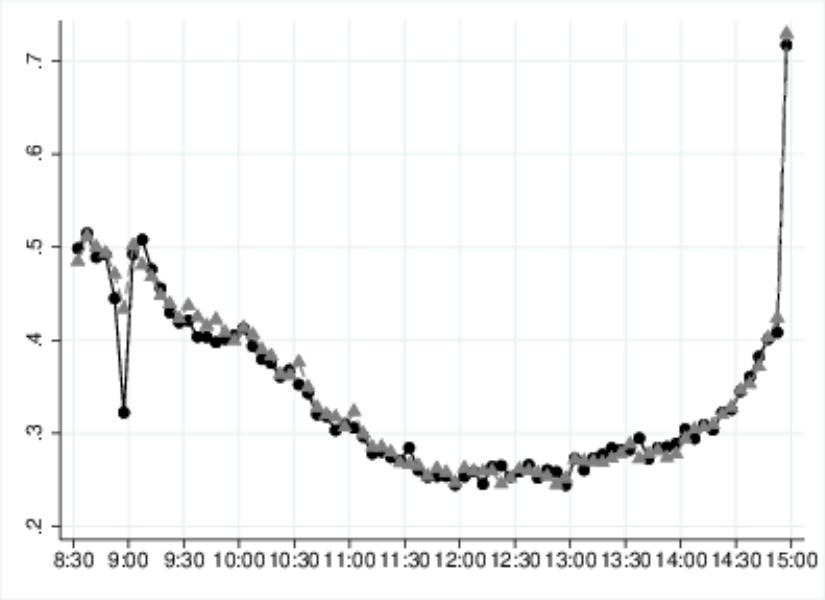} \\
$D^{-1}$ & $SPR$ \\
\includegraphics[width=.47\linewidth]{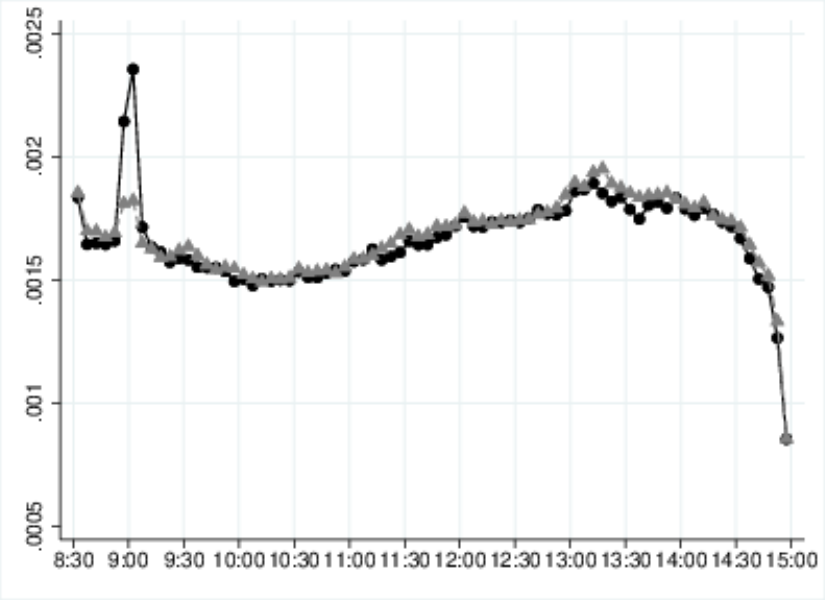} & \includegraphics[width=.47\linewidth]{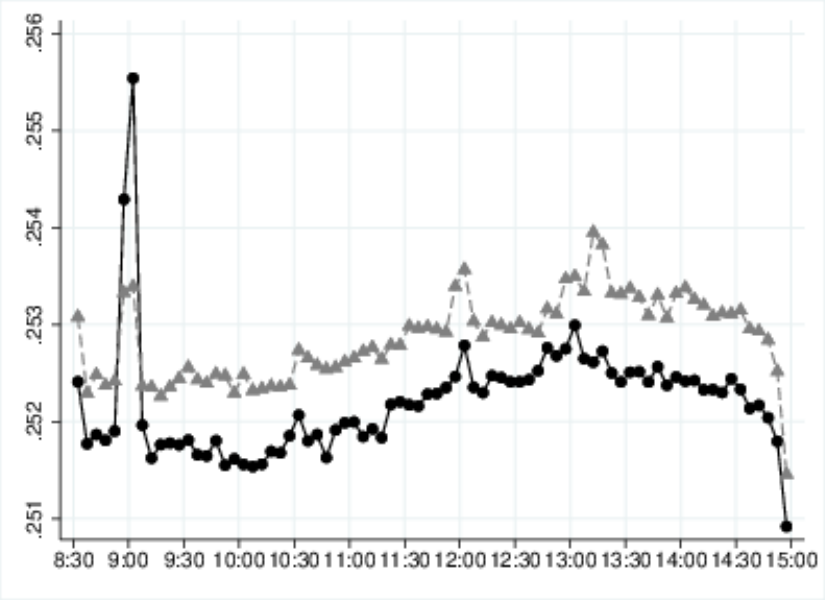} 
\end{tabular}
\caption{Average values of structural parameters ($b_{r}$, $b_{f}$, $\omega_{r}$, and $\omega_{f}$) and market activity variables ($D^{-1}$ and $SPR$) on days with macroeconomic announcements released at 9:00 (Chicago time; black circles, solid line) and on days without announcements (gray triangles, dashed line). Estimates of $b_{r}$ and $b_{f}$ are averaged over 15-minute intervals, whereas the other variables are averaged over five-minute intervals.}
\label{fig-intra-ith-par-ann}
\end{figure}

Figure~\ref{fig-intra-ith-par-ann} compares the average estimates of key structural parameters ($b_{r}$, $b_{f}$, $\omega_{r}$, and $\omega_{f}$) and market activity variables ($D^{-1}$ and $SPR$) on days with and without a 9:00 announcement. Several patterns emerge:
\begin{itemize}
\item \textbf{Price impact ($b_{r}$) rises sharply around 9:00 on announcement days.} This means that a given order flow imbalance moves prices more when news is released, consistent with liquidity deterioration reflected in higher $D^{-1}$ and wider spreads ($SPR$).
\item \textbf{Flow impact ($b_{f}$) drops around 9:00.} Order flows become less sensitive to price changes, indicating that traders pull back from price-contingent strategies before and immediately after announcements.
\item \textbf{Return innovation volatility ($\omega_{r}$) spikes just after announcements.} In the 9:00–9:05 window, $\omega_{r}$ is markedly higher, reflecting abrupt price adjustments to newly released information.
\item \textbf{Flow innovation volatility ($\omega_{f}$) declines before announcements.} In the 8:55–9:00 interval, $\omega_{f}$ falls as traders temporarily refrain from submitting orders in anticipation of news.
\end{itemize}

To formalize these patterns, we estimate regressions with two announcement dummies. The first, $ANN_{t}$, equals one if any of the eight announcements are released in interval $t$; the second, $ANN_{t}^{-}$, equals one if the announced value falls below the consensus forecast. Leads and lags of these dummies are included to capture pre- and post-announcement effects. Regressions are run separately for parameters computed over 15-minute intervals ($b_{r}$ and $b_{f}$) and over 5-minute intervals ($\omega_{r}$ and $\omega_{f}$), with controls for market activity variables—the reciprocal of depth $D^{-1}$ defined in Eq.~\eqref{eq-depth}, the number of order book events $NE$, the average size of events $ASE$, and the average spread $SPR$—as well as time dummies.

Table~\ref{tab-reg-var15m-5m-ann} presents the results. The key findings are:
\begin{itemize}
\item \textbf{Price impact ($b_{r}$):} The coefficient on $ANN_{t-1}$ is significantly positive, indicating that price impact rises just before announcements as traders withdraw liquidity. $ANN_{t}^{-}$ is also significantly positive, showing that negative news further amplifies price impact.
\item \textbf{Flow impact ($b_{f}$):} Coefficients on $ANN_{t}$ and $ANN_{t-1}$ are significantly negative, consistent with order flows becoming less responsive to prices before and after announcements. $ANN_{t}^{-}$ is likewise significantly negative, implying that negative announcements further reduce flow responsiveness.
\item \textbf{Return volatility ($\omega_{r}$):} $ANN_{t}$ is significantly positive, while $ANN_{t-2}$ and $ANN_{t+1}$ are significantly negative, indicating a volatility spike at release followed by a gradual decline.
\item \textbf{Flow volatility ($\omega_{f}$):} $ANN_{t}$ and $ANN_{t-1}$ are significantly negative, confirming that traders reduce order submissions before and around announcements.
\end{itemize}

\begin{table}[tbp]\centering
\begin{threeparttable}
\def\sym#1{\ifmmode^{#1}\else\(^{#1}\)\fi}
\caption{Regression results with announcement dummies}\label{tab-reg-var15m-5m-ann}
{\small
\begin{tabular}{lcccc}
\addlinespace
\toprule
& $b_{r}$ & $b_{f}$ & $\omega_{r}$ & $\omega_{f}$ \\
\midrule
$ANN_{t}$   &       0.025         &      -0.033\sym{**} &       0.066\sym{***}&      -0.039\sym{**} \\
            &     (0.043)         &     (0.013)         &     (0.018)         &     (0.016)         \\
\addlinespace
$ANN_{t-1}$ &       0.129\sym{***}&      -0.078\sym{***}&      -0.014         &      -0.057\sym{***}\\
            &     (0.042)         &     (0.012)         &     (0.014)         &     (0.009)            \\
\addlinespace
$ANN_{t-2}$ & & &      -0.040\sym{***}&      -0.004         \\
            & & &     (0.014)         &     (0.014)         \\
\addlinespace
$ANN_{t+1}$ & & &      -0.032\sym{**} &       0.006         \\
            & & &     (0.014)         &     (0.014)         \\
\addlinespace
$ANN_{t}^{-}$&       0.136\sym{*}  &      -0.043\sym{***}&      -0.039         &      -0.026         \\
            &     (0.070)         &     (0.016)         &     (0.026)         &     (0.025)            \\
\addlinespace
$ANN_{t+1}^{-}$& & &      -0.042\sym{**} &       0.000         \\
            & & &     (0.019)         &     (0.023)         \\
\addlinespace
$ANN_{t+2}^{-}$& & &      -0.067\sym{**} &      -0.009         \\
            & & &     (0.032)         &     (0.018)         \\
\midrule
$\bar{R}^{2}$&       0.541         &       0.269         &       0.681         &       0.510         \\
\bottomrule
\end{tabular}
}
\begin{tablenotes}
\footnotesize
\item \textit{Notes:} The dependent variables are the price impact ($b_{r}$), flow impact ($b_{f}$), return innovation volatility ($\omega_{r}$), and flow innovation volatility ($\omega_{f}$). Reported coefficients correspond to the announcement dummy ($ANN_{t}$), the negative-announcement dummy ($ANN_{t}^{-}$), and their leads and lags, when significant. Coefficients for other control variables are nearly identical to those in Table~\ref{tab-reg-var15m-5m} and are omitted for brevity. $\bar{R}^{2}$ denotes the adjusted R-squared. Robust standard errors clustered by date (1,490 clusters) are reported in parentheses. \sym{*} \(p<0.1\), \sym{**} \(p<0.05\), \sym{***} \(p<0.01\).
\end{tablenotes}
\end{threeparttable}
\end{table}

Overall, macroeconomic announcements significantly reshape the price–flow dynamic: liquidity tightens, price impact intensifies, and order submission slows. Although the announcement dummies are statistically significant, the adjusted $R^2$ values change little when they are included, suggesting that much of the observed variation is already captured by market activity variables reflecting public information.

\subsection{Structural Parameters}\label{sec:structural-par}
The reduced-form VAR model \eqref{eq:reduced-form-VAR} is estimated using returns $r_{t}$ and order flow imbalances $f_{t}$ over one-second intervals, separately for each 15-minute interval and each day. This results in 38,740 estimations.\footnote{The number of lags $p$ chosen by the Akaike information criterion is positive in 38,725 intervals. On average, lagged returns explain 8.7\% of the variation in returns and lagged flows explain 4.6\% of the variation in flows, with $R^2$ values occasionally exceeding 20\%. These findings confirm strong auto- and cross-correlation structures, motivating the SVAR framework.}  
The variance–covariance matrices of the residuals are then used to check the rank condition in Eq.~\eqref{eq:rank}, leaving 37,029 intervals suitable for ITH estimation.

For each valid interval, the ITH method is applied to estimate the structural parameters in Eq.~\eqref{eq-B-Omega}. The price and flow impacts ($b_{r}, b_{f}$) are assumed constant within a 15-minute window, while the standard deviations of return and flow innovations ($\omega_{r}, \omega_{f}$) vary across three nested five-minute subintervals. This yields one set of $(b_{r}, b_{f})$ and three sets of $(\omega_{r}, \omega_{f})$ for each 15-minute block. Because three heteroskedastic states are used, the system is over-identified.

Table~\ref{tab-sum-ith} summarizes these results. The estimated price impact $b_{r}$ exhibits substantial variation, from nearly zero to almost five, and is statistically significant in 64\% of intervals. The mean value (0.834) is considerably larger than the conditional impact of 0.474 reported by \citet{Deuskar2011}, who used one-minute data. This suggests that shorter sampling intervals capture stronger contemporaneous impacts of order flow on prices.  

By contrast, the flow impact $b_{f}$ is more moderate, typically between zero and one, but remains significant in 71\% of intervals. The average value (0.301) is lower than the 0.55 reported by \citet{Deuskar2011}, consistent with the idea that endogeneity weakens as the measurement interval approaches the event level.  

The innovation volatilities $\omega_{r}$ and $\omega_{f}$ are highly significant in over 96\% of intervals. $\omega_{r}$ is consistently larger than $\omega_{f}$, reflecting the higher variability of returns relative to flows. These volatilities also display pronounced intraday variation, as documented earlier in Section~\ref{sec:data}.

\begin{table}[tbp]\centering
\begin{threeparttable}
\caption{Estimation results for structural parameters}\label{tab-sum-ith}
{\small
\begin{tabular}{lrrrrrrrrrc}
\addlinespace
\toprule
& \multicolumn{1}{c}{Mean} & \multicolumn{1}{c}{SD} & \multicolumn{1}{c}{1\%} & \multicolumn{1}{c}{5\%} & \multicolumn{1}{c}{25\%} & \multicolumn{1}{c}{50\%} & \multicolumn{1}{c}{75\%} & \multicolumn{1}{c}{95\%} & \multicolumn{1}{c}{99\%} & \multicolumn{1}{c}{$*$} \\
\midrule
$b_{r}$ & 0.834 & 0.999 & 0.000 & 0.000 & 0.242 & 0.556 & 1.065 & 2.576 & 4.950 & 0.64 \\
\addlinespace
$b_{f}$ & 0.301 & 0.242 & 0.000 & 0.000 & 0.133 & 0.246 & 0.419 & 0.772 & 1.088 & 0.71 \\
\addlinespace
$\omega_{r,1}$ & 0.574 & 0.408 & 0.123 & 0.198 & 0.335 & 0.475 & 0.682 & 1.282 & 2.135 & 0.98 \\
\addlinespace
$\omega_{r,2}$ & 0.554 & 0.382 & 0.121 & 0.191 & 0.326 & 0.463 & 0.659 & 1.227 & 2.049 & 0.98 \\
\addlinespace
$\omega_{r,3}$ & 0.552 & 0.393 & 0.122 & 0.192 & 0.324 & 0.459 & 0.654 & 1.227 & 2.069 & 0.98 \\
\addlinespace
$\omega_{f,1}$ & 0.344 & 0.242 & 0.068 & 0.105 & 0.188 & 0.280 & 0.424 & 0.814 & 1.223 & 0.96 \\
\addlinespace
$\omega_{f,2}$ & 0.339 & 0.234 & 0.068 & 0.104 & 0.186 & 0.278 & 0.419 & 0.788 & 1.228 & 0.96 \\
\addlinespace
$\omega_{f,3}$ & 0.344 & 0.245 & 0.066 & 0.102 & 0.185 & 0.278 & 0.425 & 0.814 & 1.252 & 0.96 \\
\bottomrule
\end{tabular}
}
\begin{tablenotes}
\footnotesize
\item \textit{Notes:} For each 15-minute interval satisfying the rank condition in Eq.~\eqref{eq:rank}, the price and flow impacts ($b_{r}$ and $b_{f}$) and three sets of innovation volatilities ($\omega_{r,1}$, $\omega_{r,2}$, $\omega_{r,3}$, $\omega_{f,1}$, $\omega_{f,2}$, and $\omega_{f,3}$) are estimated using the ITH method, yielding 37,029 estimates in total. The last column ($*$) reports the proportion of intervals in which the coefficient is statistically significant, i.e., the absolute $t$-value exceeds two.
\end{tablenotes}
\end{threeparttable}
\end{table}

\subsection{Impulse Responses}\label{sec:irf}
\begin{figure}[tbp]
\centering
\begin{tabular}{cc}
Return-to-Return Impact ($IRF_{rr})$ & Flow-to-Return Impact ($IRF_{rf})$ \\
\includegraphics[width=.47\textwidth]{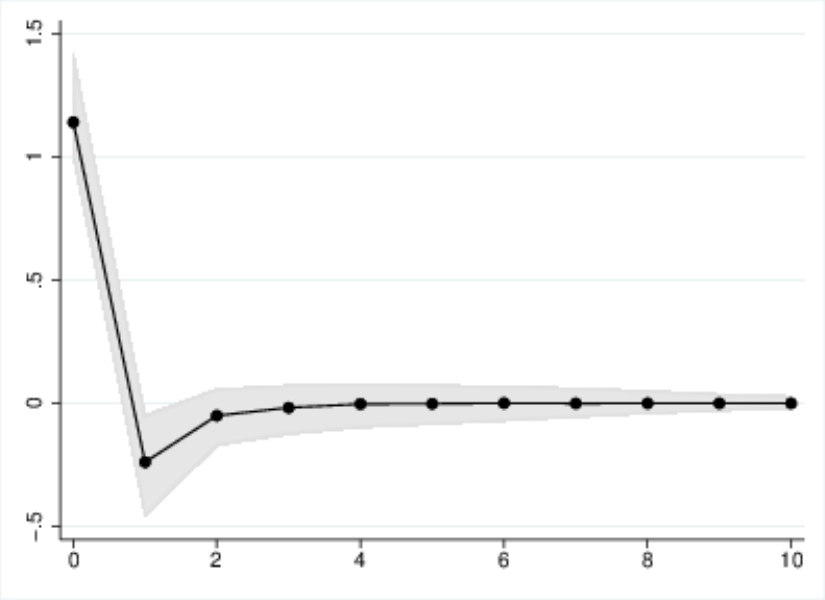} & \includegraphics[width=.47\textwidth]{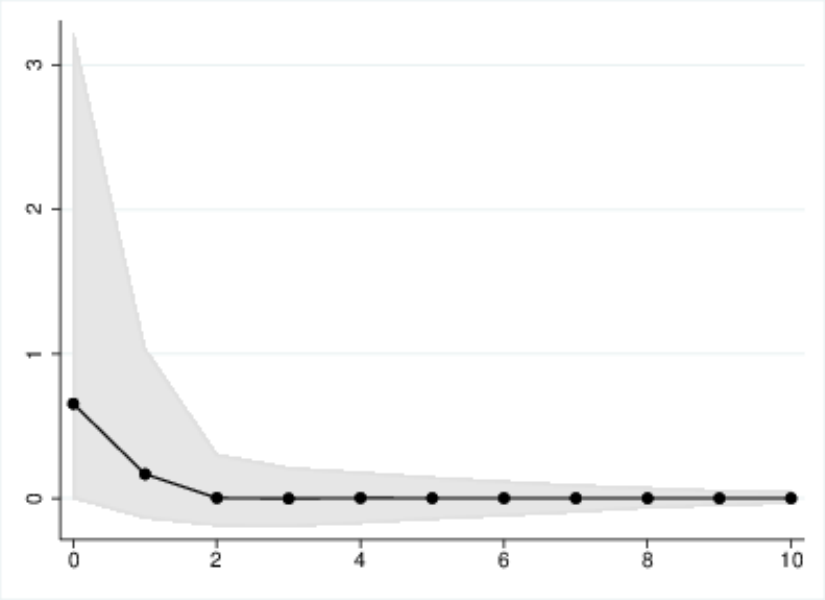} \\
Return-to-Flow Impact ($IRF_{fr})$ & Flow-to-Flow Impact ($IRF_{ff})$ \\
\includegraphics[width=.47\textwidth]{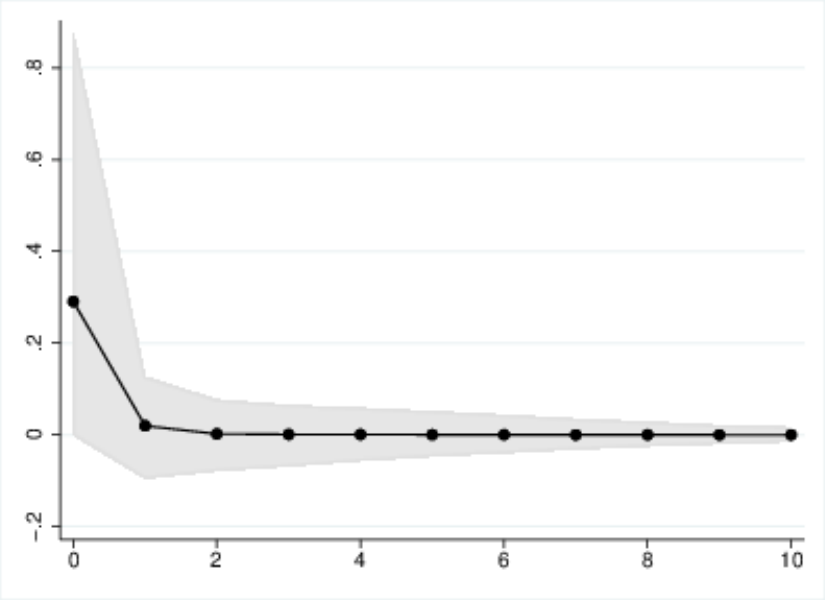} & \includegraphics[width=.47\textwidth]{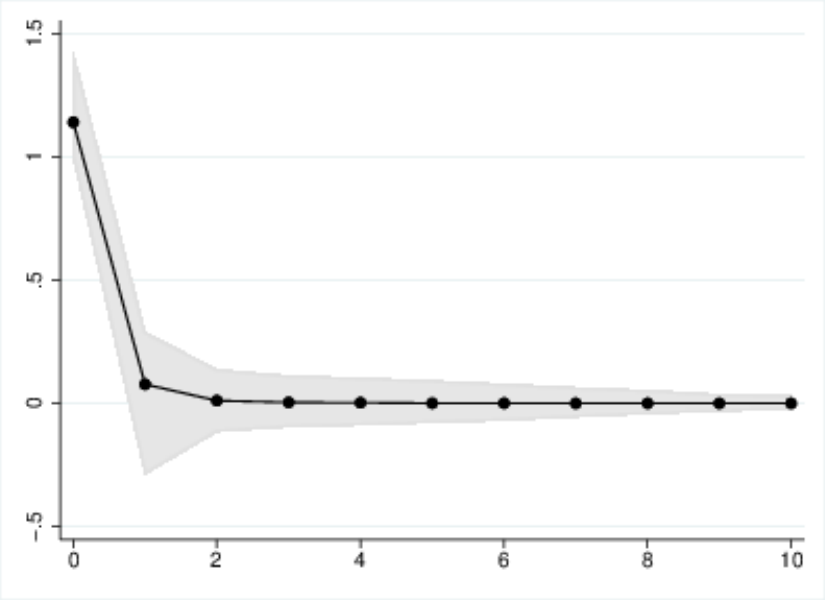}
\end{tabular}
\caption{Impulse responses to a one-unit shock in return and flow innovations, up to 10 lags. $IRF_{ij}$ denotes the response of variable $i$ to a shock in variable $j$. The figure shows the median of the 37,029 estimated impulse responses. The shaded area indicates the 5th and 95th percentiles.}
\label{fig-irf}
\end{figure}

Having identified the structural parameters, we now analyze the dynamic interactions between returns and flows. Figure~\ref{fig-irf} shows the median impulse responses of returns and flows to shocks, along with 5th and 95th percentiles, computed from the 37,029 estimations.  

Several patterns emerge. First, most impacts dissipate quickly, with responses beyond one lag nearly negligible. This implies that market adjustments to order flow and return shocks occur within a second, consistent with \citet{Hautsch2012a}, who report that quotes reach new levels after roughly 20 events—less than the average number of events in one second in our sample.  

The instantaneous responses $IRF_{ij}(0)$ are almost uniformly positive, confirming significant contemporaneous interactions between return and flow shocks. The first-lag return-to-return response $IRF_{rr}(1)$ is negative in over 95\% of cases, suggesting short-term reversal in returns. Other first-lag responses fluctuate around zero with no consistent sign.  

Table~\ref{tab-sum-iinf} reports summary statistics of the long-run impacts $I_{ij}(\infty)$ defined in Eq.~\eqref{eq-irf-inf}. All are positive on average except for $I_{fr}(\infty)$ in some intervals. A one-unit flow shock, corresponding to an unexpected increase of about 1,000 contracts at the best bid or ask, raises prices by roughly 1.42 basis points in the long run. Using the mean flow volatility $\omega_{f}$ (0.34, about half the mean depth in Table~\ref{tab-sumstat}), this translates into a price impact of 0.48 basis points—closely aligned with the 0.5–0.6 basis points reported by \citet{Hautsch2012a} for limit orders of half-depth.  

Taken together, these results show that while return–flow interactions are strong at the contemporaneous level, they dissipate extremely rapidly. The long-run effects of flow shocks are economically meaningful and consistent with established findings in the market microstructure literature.

\begin{table}[tbp]\centering
\begin{threeparttable}
\caption{Summary statistics of long-run impacts}\label{tab-sum-iinf}
{\small
\begin{tabular}{lrrrrrrrrr}
\addlinespace
\toprule
& \multicolumn{1}{c}{Mean} & \multicolumn{1}{c}{SD} & \multicolumn{1}{c}{1\%} & \multicolumn{1}{c}{5\%} & \multicolumn{1}{c}{25\%} & \multicolumn{1}{c}{50\%} & \multicolumn{1}{c}{75\%} & \multicolumn{1}{c}{95\%} & \multicolumn{1}{c}{99\%} \\
\midrule
$I_{rr}(\infty)$ & 0.843 & 0.283 & 0.338 & 0.499 & 0.702 & 0.834 & 0.980 & 1.220 & 1.399 \\
\addlinespace
$I_{rf}(\infty)$ & 1.423 & 16.688 & 0.038 & 0.180 & 0.454 & 0.817 & 1.580 & 4.122 & 8.464 \\
\addlinespace
$I_{fr}(\infty)$ & 0.372 & 0.721 & -0.212 & -0.037 & 0.171 & 0.320 & 0.523 & 0.977 & 1.455 \\
\addlinespace
$I_{ff}(\infty)$ & 1.304 & 6.925 & 0.514 & 0.735 & 1.056 & 1.224 & 1.448 & 1.861 & 2.226 \\
\bottomrule
\end{tabular}
}
\begin{tablenotes}
\footnotesize
\item \textit{Notes:} Summary statistics are based on 37,029 estimates of the long-run impacts $I_{ij}(\infty)$ defined in Eq.~\eqref{eq-irf-inf}.
\end{tablenotes}
\end{threeparttable}
\end{table}

\subsection{Intraday Variations}
\label{sec:intraday-variations}

\begin{figure}[tbp]
\centering
\begin{tabular}{cc}
$b_{r}$ & $b_{f}$ \\
\includegraphics[width=.47\linewidth]{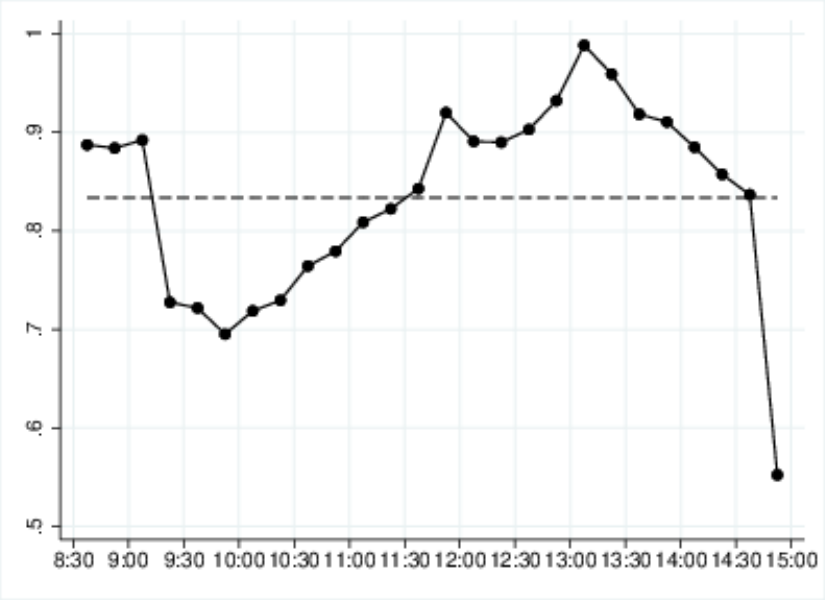} & \includegraphics[width=.47\linewidth]{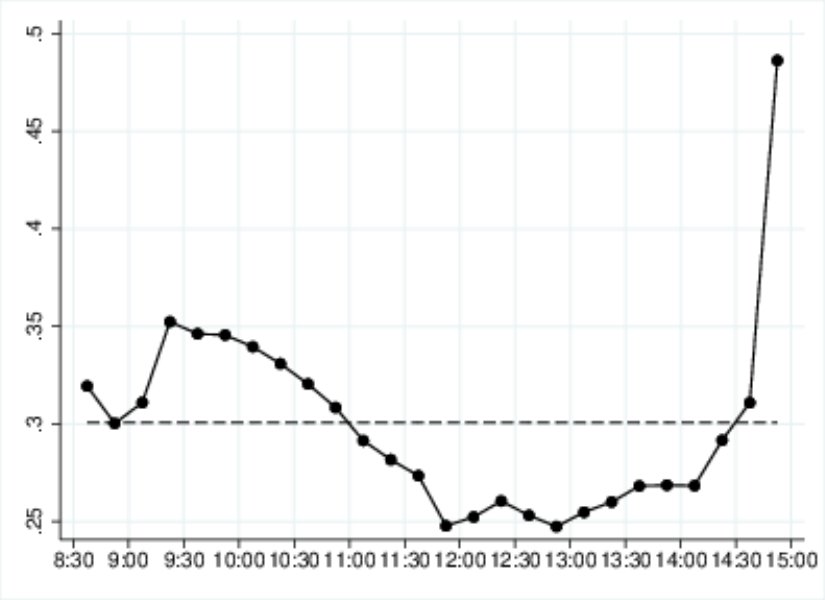} \\
$\omega_{r}$ & $\omega_{f}$ \\
\includegraphics[width=.47\linewidth]{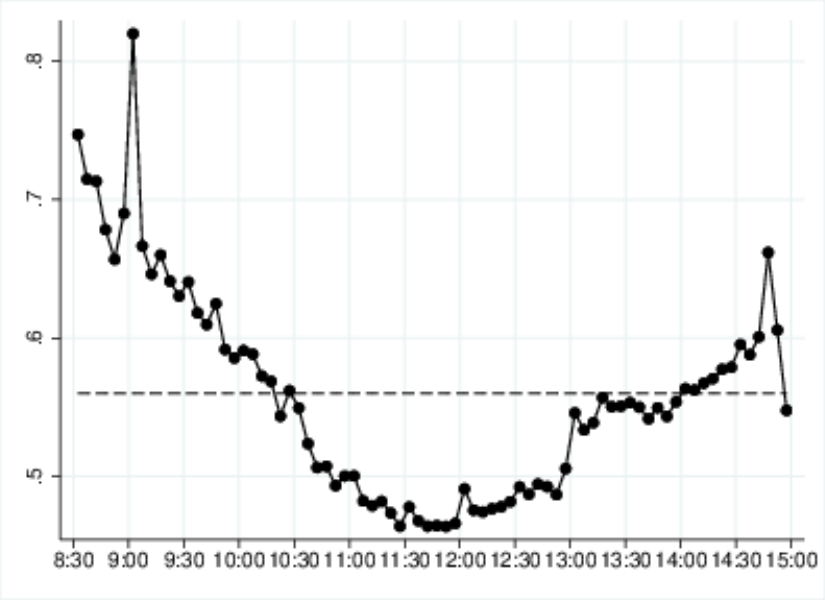} & \includegraphics[width=.47\linewidth]{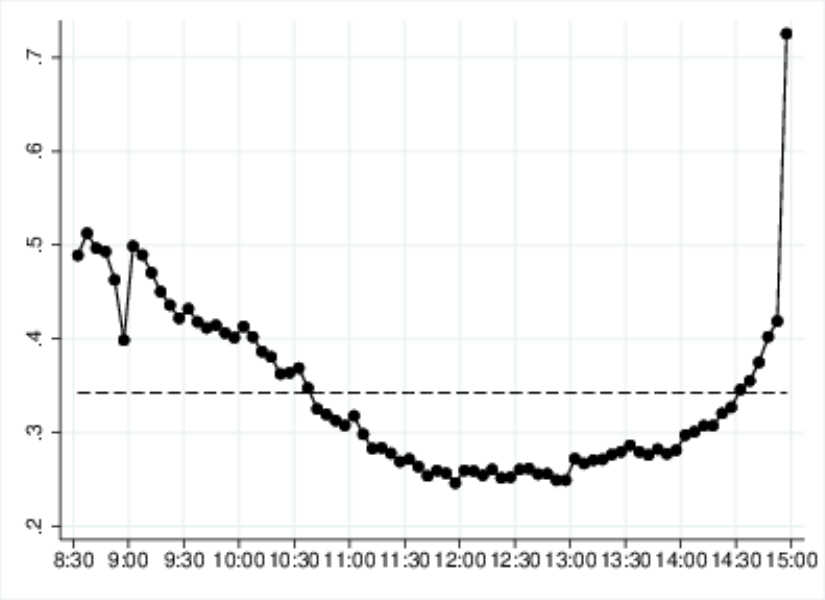}
\end{tabular}
\caption{Intraday averages of estimated price and flow impacts ($b_{r}$ and $b_{f}$) and standard deviations of return and flow innovations ($\omega_{r}$ and $\omega_{f}$). Averages of $b_{r}$ and $b_{f}$ are computed for each 15-minute interval, whereas those of $\omega_{r}$ and $\omega_{f}$ are computed for each five-minute interval. Dashed lines indicate the full-sample averages.}
\label{fig-intra-ith-par}
\end{figure}

We next examine how the structural parameters and innovation volatilities evolve within the trading day. Figure~\ref{fig-intra-ith-par} plots the averages of $b_{r}$, $b_{f}$, $\omega_{r}$, and $\omega_{f}$ across 15-minute or 5-minute intervals, respectively.

The estimated price impact $b_{r}$ is elevated in the opening interval (8:30--9:15), drops around 9:15--9:30, and then gradually rises over the trading session, peaking around 13:00--13:15 before declining sharply into the close. By contrast, the flow impact $b_{f}$ increases in 9:00--9:30, declines steadily until midday, reaches its lowest point around 12:45--13:00, and then surges in the final interval (14:45--15:00). This pattern suggests that returns are least sensitive to flows in the late afternoon, while flows become increasingly driven by returns toward the close, consistent with heightened liquidity and trading activity during that period (see Figure~\ref{fig-intra-var}).

The standard deviations of return and flow innovations, $\omega_{r}$ and $\omega_{f}$, mirror the volatility dynamics of returns and order flow imbalances, respectively. Notably, $\omega_{r}$ spikes in 9:00--9:05, while $\omega_{f}$ falls in 8:55--9:00, patterns linked to macroeconomic announcements as shown in Section~\ref{sec:macro-news}. In addition, $\omega_{f}$ rises sharply in the last five minutes (14:55--15:00), reflecting aggressive order submission near the close.

To investigate the drivers of these patterns, we regress the estimated parameters on market activity measures. Table~\ref{tab-reg-var15m-5m} reports the results. For $b_{r}$, the coefficient on reciprocal depth ($D^{-1}$) is strongly positive, confirming that order flows exert larger price impacts when the market is thin. This estimate is close to the theoretical benchmark implied by the stylized LOB model of \citet{Cont2014}. By contrast, $NE$ (number of events) and $ASE$ (average size of events) both enter negatively, indicating that more active and aggressive order submissions reduce the contemporaneous price impact. A wider spread ($SPR$) also lowers $b_{r}$. Together with time dummies, these variables explain about 54\% of the variation in $b_{r}$.

For $b_{f}$, the coefficients on $D^{-1}$ and $SPR$ are significantly negative and positive, respectively, implying that returns induce more order flow imbalances when the market is thin or trading costs are high. $ASE$ enters positively, showing that flows are more sensitive to returns when order submissions are aggressive. The explanatory power (27\%) is about half that for $b_{r}$, consistent with the idea that additional factors govern return-induced flows.

Turning to innovation volatilities, $\omega_{r}$ rises with illiquidity ($D^{-1}$), while $\omega_{f}$ falls, indicating that thin markets amplify return shocks but dampen flow shocks. Both measures increase with more active ($NE$) and aggressive ($ASE$) order submissions, as well as with wider spreads. These variables account for 68\% and 51\% of the variation in $\omega_{r}$ and $\omega_{f}$, respectively.

\begin{table}[tbp]\centering
\def\sym#1{\ifmmode^{#1}\else\(^{#1}\)\fi}
\begin{threeparttable}
\caption{Regression results for intraday variations}\label{tab-reg-var15m-5m}
{\small
\begin{tabular}{lcccc}
\addlinespace
\toprule
& $b_{r}$ & $b_{f}$ & $\omega_{r}$ & $\omega_{f}$ \\
\midrule
$D^{-1}$ & 0.553\sym{***} & -0.066\sym{***} & 0.204\sym{***} & -0.049\sym{***} \\
& (0.017) & (0.004) & (0.007) & (0.002) \\
\addlinespace
$NE$ & -1.628\sym{***} & 0.009 & 12.357\sym{***} & 9.423\sym{***} \\
& (0.429) & (0.095) & (0.434) & (0.449) \\
\addlinespace
$ASE$ & -3.421\sym{***} & 1.502\sym{***} & 1.799\sym{***} & 4.471\sym{***} \\
& (0.834) & (0.266) & (0.279) & (0.640) \\
\addlinespace
$SPR$ & -7.510\sym{***} & 3.173\sym{***} & 7.302\sym{***} & 8.833\sym{***} \\
& (2.868) & (0.413) & (0.732) & (0.914) \\
\midrule
$\bar{R}^{2}$ & 0.541 & 0.268 & 0.681 & 0.509 \\
\bottomrule
\end{tabular}
}
\begin{tablenotes}
\footnotesize
\item \textit{Notes:} The dependent variables are structural parameters ($b_{r}$ and $b_{f}$) estimated over 15-minute intervals, and innovation volatilities ($\omega_{r}$ and $\omega_{f}$) estimated over five-minute intervals. The explanatory variables include reciprocal depth $D^{-1}$ (per thousand), number of events $NE$ (in millions), average event size $ASE$ (in thousands), and average spread $SPR$, along with time dummies. Robust standard errors clustered by date (1,490 clusters) are reported in parentheses. \sym{*} \(p<0.1\), \sym{**} \(p<0.05\), \sym{***} \(p<0.01\).
\end{tablenotes}
\end{threeparttable}
\end{table}

Overall, these results show that intraday fluctuations in structural parameters are closely tied to liquidity and trading activity. Combined with the announcement effects documented in Section~\ref{sec:macro-news}, they underscore how both public information releases and market microstructure conditions shape the dynamics between order flows and returns throughout the trading day.

\section{Conclusion}
\label{sec:conclusion}

This paper examines the interaction between returns and order flow imbalances in the S\&P 500 E-mini futures market using a structural VAR identified through heteroskedasticity. By estimating the model at the one-second frequency and separately for each 15-minute interval, we capture both intraday variation and the endogeneity between prices and flows.

The empirical analysis yields several insights. Macroeconomic news announcements sharply alter high-frequency dynamics: price impact rises, flow impact declines, return volatility spikes, and flow volatility falls, reflecting temporary liquidity withdrawal. Beyond announcements, structural linkages between returns and flows remain economically significant at the one-second horizon. Price impact estimates are consistent with stylized LOB models, while flow impact confirms feedback from prices to order submissions. Impulse response analysis shows that shocks dissipate within a second, highlighting the efficiency of high-frequency price discovery. Finally, strong intraday patterns in impacts and volatilities are closely tied to market activity measures such as depth, order frequency, and spreads.

Overall, the results underscore how liquidity provision and order submission adapt to scheduled announcements and intraday conditions, offering new evidence on the microstructure of modern electronic markets. Future work could extend the framework to multivariate settings that distinguish order types or explore cross-market interactions between related assets.

\bibliographystyle{apa}
\bibliography{svar2023}

\end{document}